# Scaling laws of multi-shock implosions toward the quasi-isentropic limit

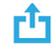

M. Murakami[a]

**AFFILIATIONS**

Institute of Laser Engineering, Osaka University, Suita, Osaka 565-0871, Japan

[a]Author to whom correspondence should be addressed: murakami.masakatsu.ile@osaka-u.ac.jp

**ABSTRACT**

We present a unified theoretical and numerical framework for self-similar multi-shock implosions achieving ultra-high compression in a uniform solid spherical target. Extending the classical Guderley model to $N$-stacked, spherically converging shocks, we derive self-similar solutions and the scaling law for the final density of the form $\rho_r/\rho_0 \propto \hat{P}^{\beta(N-1)}$, where $\hat{P}$ is the stage pressure ratio and $\beta$ is determined by the adiabatic index $\gamma$. One-dimensional Lagrangian hydrodynamic simulations confirm this relation over a broad range of parameters, from the weakly to the strongly nonlinear regime ($\hat{P} \sim 70$). The results show that cumulative compression increases systematically with the number of stacked shocks while entropy generation is strongly suppressed, asymptotically approaching a quasi-isentropic limit as $N \to \infty$. This volumetric scheme strongly suppresses the Rayleigh–Taylor instability that plagues shell-based implosions and thus provides a robust, largely instability-resistant compression pathway applicable to inertial confinement fusion and other high-energy-density systems. The framework bridges similarity theory with realistic multi-shock dynamics, guiding the design of advanced laser-driven compression schemes.

## I. INTRODUCTION

Self-similar solutions have long served as a cornerstone of theoretical hydrodynamics, providing analytic insight into extreme compression and flow focusing. A landmark example is the Guderley solution,[1–3] which describes a single converging shock wave focused toward the origin in a spherically symmetric compressible fluid. Independently derived by Stanyukovich[4] and later extended by Gandel'man and Frank-Kamenetskii[5] to include the reflected-shock stagnation stage, this class of similarity solutions has provided deep insights into inertial-fusion implosions,[6–8] astrophysical collapse, and high-energy-density plasma physics. Subsequent studies (e.g., Refs. 9–11) further explored related multi-shock or reflection structures. The present work, however, focuses on a unified self-similar framework for stacked converging shocks in solid-density matter.[12]

Recent advances in laser compression and high-energy-density diagnostics call for rigorous theoretical models that go beyond empirical optimization and offer predictive capability under idealized yet physically transparent assumptions. While extensive studies have addressed single-shock or ablatively driven similarity flows, no analytic framework has yet been developed for self-similar multi-shock implosions in solid-density matter. This paper presents a systematic formulation of such solutions, constructing similarity flows for a stacked sequence of converging shocks in which each shock successively compresses the core with increasing intensity.

Although developed within ideal hydrodynamics, the model is directly motivated by applications in inertial confinement fusion (ICF), where maximizing central compression while maintaining hydrodynamic stability is essential. Conventional thin-shell implosions are inherently prone to Rayleigh–Taylor (R–T) instability during both acceleration and deceleration phases. In contrast, the shock-focused geometry analyzed here delivers energy volumetrically without shell ablation, naturally mitigating R–T growth and reducing the need for elaborate stabilization schemes. This volumetric compression mechanism provides a physically robust pathway toward stable and efficient implosions.

By focusing multiple converging shocks toward the center in a temporally staggered manner, significantly higher compressions can be achieved than with a single shock. Such stacked-shock schemes, therefore, offer a potential foundation for advanced ignition concepts.

The ideal Guderley solution predicts diverging pressure and temperature at the point of shock reflection. In realistic systems, however, this singular behavior is mitigated by dissipative effects such as electron thermal conduction. A previous study by Murakami et al.[13] showed that thermal diffusion leads to finite stagnation temperatures even for strongly imploding shocks. These effects will be examined in greater detail in a forthcoming companion paper.

Toward the goal of achieving extreme compression in laser-driven systems, the concept of multiple shock convergence has already been employed in modern ICF experiments, most notably at the National Ignition Facility (NIF).[14,15] There, multi-step shock delivery is realized through engineered laser pulse shaping to control shock timing and coalescence.[16] Related concepts, such as shock ignition,[17] further illustrate the importance of controlled multi-shock dynamics in high-energy-density implosions.

It is important to note that the requirement of a stacked (or staged) shock structure in convergent compression is not specific to modern inertial fusion concepts.[26]

Such experimental configurations, however, remain largely empirical, relying on extensive simulations and fine-tuned design parameters, without invoking an analytic framework for the underlying similarity structure or limiting behavior. In contrast, this work presents a systematic theory based on self-similar solutions for $N$-stacked converging shocks, thereby extending the classical Guderley model into a multi-shock regime.

This framework enables the derivation of scaling laws, density amplification factors, and entropy profiles under idealized conditions, providing both fundamental physical insight and a benchmark for assessing the ultimate performance limits of experimentally realizable implosion schemes.

The paper is organized as follows. Section II outlines the self-similar theory of stacked converging shocks and its relation to the single-shock Guderley solution. Section III develops compression-scaling laws, analyzes areal density and entropy reduction, and compares with simulations. Section IV discusses structural advantages, dissipation, and relevance to ignition. The *Summary* concludes with key results and outlook.

From a fundamental viewpoint, however, the necessity of stacked shocks is not merely empirical. As demonstrated in Appendix A, it follows directly from the causal structure of spherical compressible flows: a single strong converging shock inevitably breaks causal control and produces excessive entropy, whereas a properly staged sequence of shocks preserves causal connectivity and enables controlled high compression.

Part of the theoretical framework and numerical results presented in this work build upon our recent publication in Phys. Rev. E,[20] which focused on self-similar multishock implosions driven by a small number of strong converging shocks. In contrast, the present study substantially extends this earlier work by investigating sequences of multiple weak shocks and their cumulative compression effects, particularly relevant for inertial confinement fusion. We show that the total compression can be preserved while the entropy production is significantly reduced as the number of stacked shocks increases.

## II. SELF-SIMILAR SOLUTION FOR STACKED SHOCK WAVES

Throughout this study, we focus on the essential hydrodynamic features by omitting laser–matter interactions, heat conduction, and radiation transport, and by employing an ideal equation of state. Figure 1 illustrates the temporally stacked pressure sequence ($P_1 - P_4$) used to generate successive shocks coalescing at the center. The initial pressure of the target material, $p_0$, is assumed to be much lower than $P_1$ ($p_0 \ll P_1$).

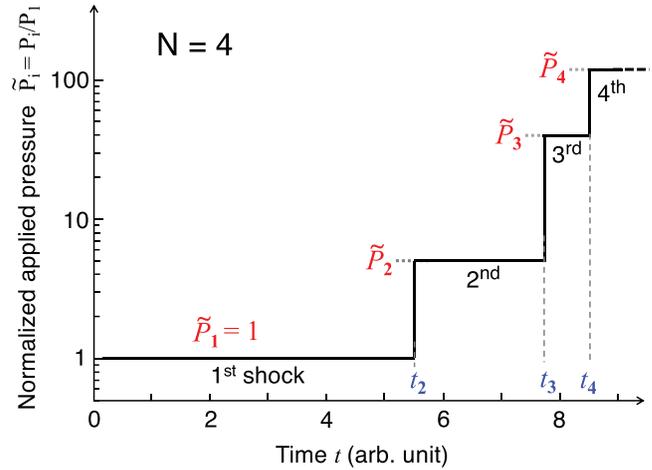

**FIG. 1.** Illustrative example of a time-dependent external pressure sequence used to drive a multi-shock implosion (as a representative but non-optimized configuration). A series of four shocks is generated by applying stepwise increasing pressures ($P_1$, $P_2$, $P_3$, $P_4$), each precisely timed ($t_2$, $t_3$, $t_4$) to ensure their simultaneous convergence at the center.

### A. 1D hydrodynamic simulations demonstrating self-similarity

In this subsection, we present one-dimensional (1D) hydrodynamic simulations of spherical compression in a uniform solid target composed of a deuterium–tritium (DT) mixture. It is well known that a single spherically converging shock wave propagating through a uniform medium evolves asymptotically toward a self-similar state in pressure, density, and velocity near the moment of collapse.[2] Our aim here is to demonstrate that the self-similar behavior also emerges in the case of multiple stacked shocks.

Figure 2(a) shows a flow diagram obtained from the 1D hydrodynamic simulation. The blue curves represent Lagrangian mesh trajectories. The target is a uniform solid DT sphere with an initial density of $\rho_0 = 0.2$ g/cm$^3$ and an initial radius of $R_0 = 1$ mm. The implosion is driven by two sequential external pressures applied to the outer surface of the target. The first pressure, $P_1 = 10$ Mbar, is applied for $t \leq 6$ ns, followed by a second pressure, $P_2 = 80$ Mbar, applied for $t \geq 6$ ns. These pressures and their timing are optimized such that the resulting two shocks converge and collide simultaneously at the center.

Figure 2(b) presents an enlarged view of the region highlighted by the red square in Fig. 2(a), illustrating the final stage of the simultaneous shock collapse at $t = t_c$. The three red curves trace the shock front trajectories, each of which follows the form $r \sim |t - t_c|^\alpha$. Here $\alpha$ denotes the similarity exponent characterizing the temporal scaling of the shock-front radius near collapse. While the extended red curves are also visible at larger radii in Fig. 2(a), they show increasing deviations from the corresponding shock fronts obtained in the simulation. This is because the self-similar behavior becomes more pronounced at smaller radii. The first converging shock follows the similarity exponent $\alpha = \alpha^* = 0.688\,38$, where $\alpha^*$ denotes the classical Guderley exponent for a single converging shock (hereafter we use the asterisk "$*$" to denote the $N = 1$ case). The second converging shock and the reflected (diverging) shock follow a slightly different exponent $\alpha = \alpha_1 = 0.693\,45$, as obtained from the multi-stacked self-similar solution described later in more detail. At a glance, however, the curves corresponding to $\alpha^*$ and $\alpha_1$ appear nearly identical.

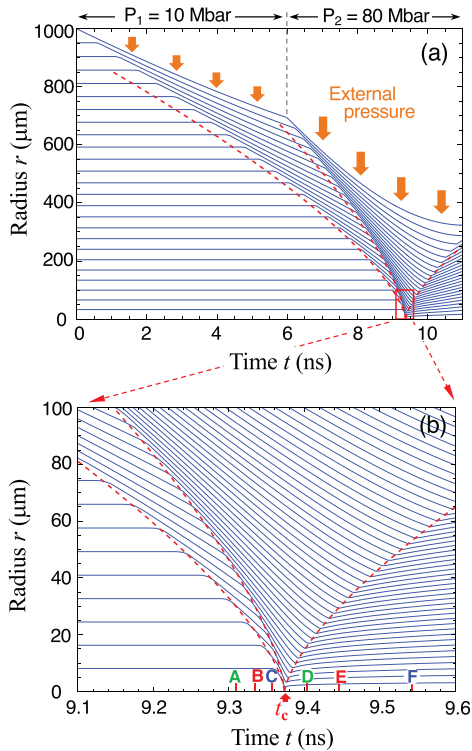

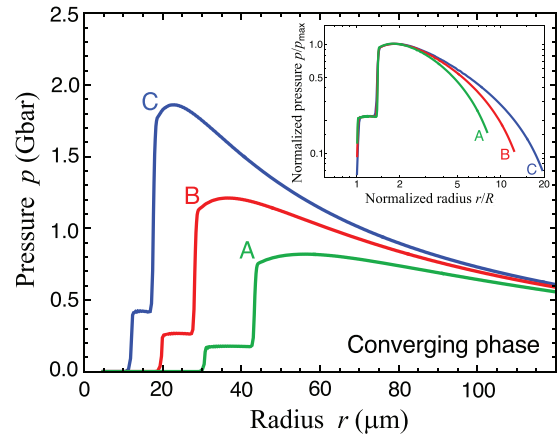

FIG. 2. (a) Lagrangian mesh trajectories (blue curves) demonstrate the implosion dynamics of a solid DT sphere subjected to two-stage pressure loading: 10 Mbar for $t \leq 6$ ns, followed by 80 Mbar for $t \geq 6$ ns. The pressure timing is optimized so that the resulting two converging shock fronts simultaneously collide at the center. (b) Enlarged view of the region highlighted by the red square in panel (a), illustrating the final stage of the simultaneous shock collapse at $t = t_c$. The shock front trajectories follow a self-similar scaling of the form $r \propto |t - t_c|^\alpha$. Reproduced with permission from Murakami, Phys. Rev. E **112**, 055206 (2025) under CC BY 4.0.[20]

FIG. 3. Self-similar evolution of pressure profiles during convergence. The figure shows pressure profiles at three successive times (A, B, and C), as marked in Fig. 2(b), during the converging phase. The peak pressure increases and the shock width narrows progressively, while the two-step shock structure is preserved. Inset: The same profiles are replotted in a log–log scale, normalized by the position of the first shock front $R(t)$ and peak pressure $p_{max}$ at each time. The excellent overlap among the curves confirms the self-similar nature of the shock evolution. Reproduced with permission from Murakami, Phys. Rev. E **112**, 055206 (2025) under CC BY 4.0.[20]

Figure 3 shows the temporal evolution of the pressure profiles at three sequential times, A, B, and C, during the converging phase, as marked in Fig. 2(b). The profiles clearly demonstrate that the pressure peak increases progressively while the spatial width becomes narrower as the shock front approaches the center. The self-similar nature of the system is confirmed clearly in the inset of Fig. 3, where the three pressure profiles are replotted in log–log scale. Both the radius and pressure are normalized by the position of the first shock front and the peak pressure at each respective time. The resulting three curves show excellent overlap, providing strong evidence for the self-similarity of the system.

### B. Review of the Guderley solution for a single converging shock

A canonical example of dynamically driven spherical compression is provided by the Guderley solution,[1] which describes a self-similar converging shock focusing at the origin in a finite time. Under spherical symmetry and a barotropic equation of state, Guderley introduced the remarkable self-similar ansatz in which the shock trajectory $R(t)$ follows a power-law form,

$$R(t) \propto |t - t_c|^\alpha, \quad (1)$$

where the exponent $\alpha$ is not imposed by the boundary drive but emerges as an eigenvalue of the self-similar problem, determined by the equation of state and regularity conditions across the converging shock (see Ref. 2 for details).

A key feature of the Guderley model is that the temporal evolution of the compression is governed by inward-propagating compressive characteristics, which establish a finite-time causal connection between the shock front and the central region. As a result, the pressure and density exhibit universal power-law behavior near the focusing time, largely independent of the details of the external drive. For brevity, the full formulation of the governing equations, the self-similar ansatz, and the associated normalized boundary conditions are not repeated here, as they are presented in detail in Ref. 21, to which the reader is referred for a complete description.

This dynamically driven self-similar collapse is fundamentally different from quasi-static isentropic compression models such as the classical Kidder model,[22] and serves as a prototypical reference for characteristic-controlled spherical implosions.

### C. Extension to multi-shock systems

With the classical Guderley solution serving as a canonical reference for dynamically driven, characteristic-controlled spherical implosions, we now extend this framework to multi-stacked shock systems. It is therefore important to clarify in what sense such an extension can still be regarded as self-similar.

Before proceeding, we briefly examine the relation between the present formulation and the earlier analysis by Lazarus.[10] Lazarus showed that exact self-similar solutions describing multiple converging shocks collapsing simultaneously in spherical geometry exist only when the adiabatic index exceeds a critical value, $\gamma > \gamma_{cr} \simeq 1.8698$.

Under this condition, all shocks share a common similarity exponent, and the corresponding integral curves pass smoothly through the singular point on a single $(U, Z)$ phase plane. Here, $(U, Z)$ denote the standard dimensionless variables representing the normalized velocity and sound-speed parameters in the self-similar phase-plane formulation. In some literature (e.g., Refs. 2 and 21) the normalized velocity is denoted by $V$ instead of $U$; in the present paper, we adopt $U$ to avoid confusion with the specific volume $V$.

In the present work, we consider the physically relevant case $\gamma = 5/3$, which lies below this critical threshold. As a result, an exact global self-similar solution connecting multiple converging shocks on a single phase plane does not exist. Instead, the flow is described by a sequence of locally self-similar solutions, each defined on its own $(U, Z)$ phase plane and characterized by a slightly different similarity exponent.

Specifically, the upstream flow of the second shock follows the classical Guderley solution with exponent $\alpha^*$, represented by the blue integral curve in Fig. 4(a). Across the second shock, the state is mapped from points 2 to 3 via the Rankine–Hugoniot relations, thereby transferring the solution onto a different phase plane. Downstream of this shock, the flow evolves according to a distinct self-similar solution characterized by the exponent $\alpha_1$, shown by the red curve in Fig. 4(a), which smoothly connects point 3 to the origin on its own sonic line.

Although the resulting structure is not strictly self-similar over the entire domain, it provides an accurate approximation to the full hydrodynamic evolution. The difference between the two similarity exponents is small, $|\alpha_1/\alpha^* - 1| \simeq 7 \times 10^{-3}$, and one-dimensional simulations confirm that the matched piecewise self-similar description reproduces the pressure and density profiles in both the converging and reflected phases (see Fig. 5). In this sense, the present formulation may be regarded as a matched asymptotic extension of the classical Guderley solution to multi-shock systems below the Lazarus critical index.

We now illustrate this extension for the two-stacked case, using the same implosion parameters as in Figs. 2 and 3. The generalization to multiple shocks is straightforward.

The initial conditions for a uniform-density spherical target are uniquely specified and correspond to point 1 in Figs. 4(a)–4(d). Before the second shock arrives, the fluid compressed by the first shock evolves exactly as in the single-shock case. Accordingly, the similarity exponent must coincide with that of the classical Guderley solution, $\alpha = \alpha^* = 0.688\,38$, ensuring that the integrated trajectory in Fig. 4(a) smoothly crosses the sonic line $Z = (U - \alpha^*)^2$. Indeed, the trajectory of the first shock shown in Fig. 2(b) follows the expected scaling law $r \propto |t - t_c|^{\alpha^*}$.

When the second shock arrives, discontinuities in density, velocity, and pressure arise as the state transitions from points 2 (upstream) to 3 (downstream), as illustrated in Fig. 4. A comparison between the Guderley solution and the 1D hydrodynamic simulation (red curves in Fig. 4) gives the upstream coordinates $(U_2, Z_2) = (0.27, 0.050)$. With the pressure jump $\tilde{p}_3 \equiv p_3/p_2 = 4.2$ obtained from Fig. 4(d), the downstream coordinates of point 3 are estimated as $(U_3, Z_3) = (0.50, 0.097)$ from the Rankine–Hugoniot jump conditions:

$$U_3 - U_2 = \left[\frac{2(\tilde{p}_3 - 1)^2 Z_2}{\gamma((\gamma + 1)\tilde{p}_3 + \gamma - 1)}\right]^{1/2}, \quad (2)$$

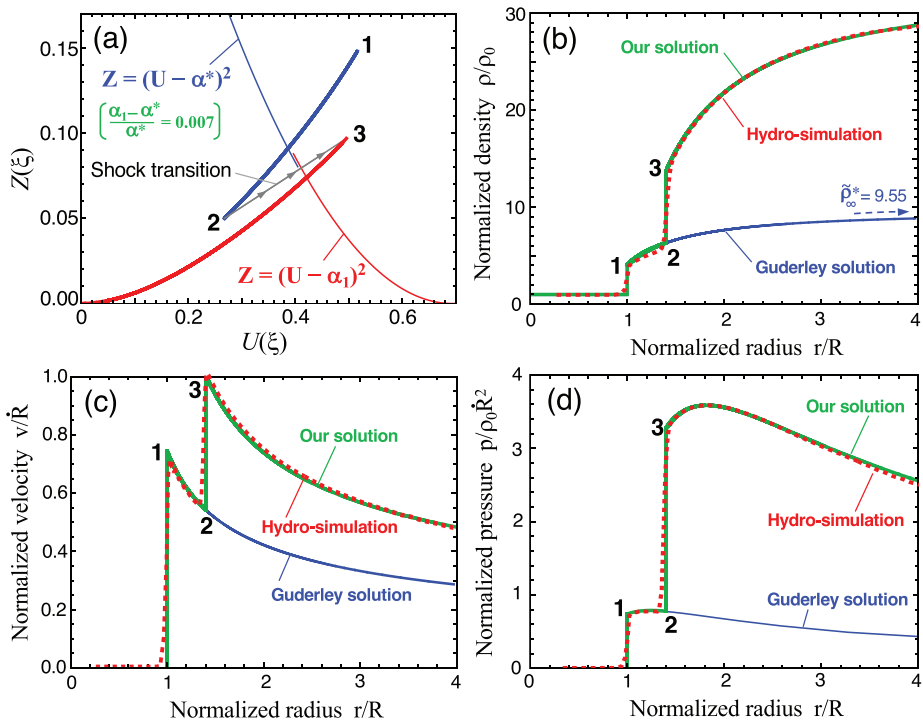

FIG. 4. (a) Integrated trajectories in the $U$–$Z$ plane for $\gamma = 5/3$, showing the first-shock (blue, $\alpha^* = 0.688\,38$) and second-shock (red, $\alpha = 0.693\,45$) solutions. (b)–(d) Scaled radial profiles of density, velocity, and pressure for multi-stacked converging shock systems discussed in Sec. II C. Solid curves: hydrodynamic simulations; dashed curves: self-similar solutions. Reproduced with permission from Murakami, Phys. Rev. E **112**, 055206 (2025) under CC BY 4.0.[20]

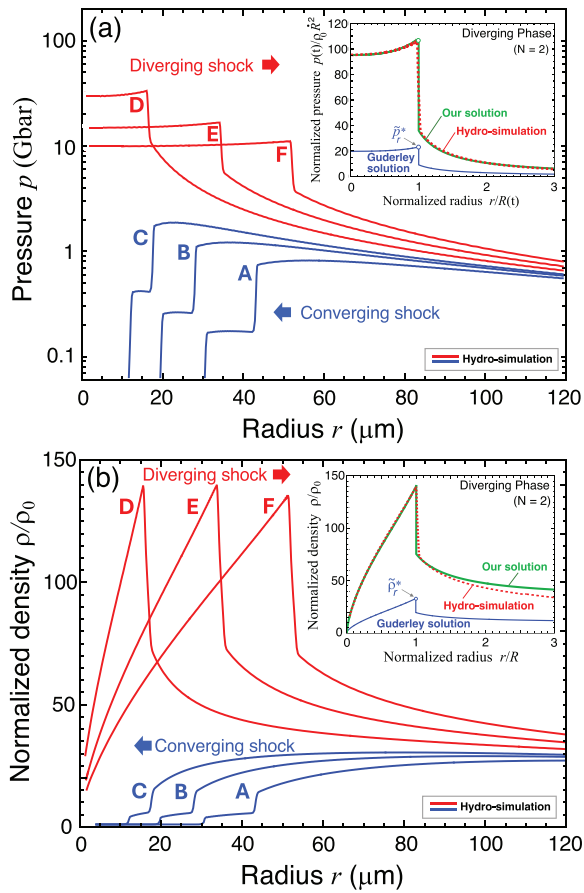

**FIG. 5.** Temporal evolution of (a) pressure and (b) density obtained from the 1D hydrodynamic simulation shown in Fig. 2. Labels A–C and D–F denote the converging and diverging phases, respectively. Insets compare the simulation results with both the two-shock self-similar solution and Guderley's original solution in the diverging phase, where the two shocks merge into a single outgoing shock. The reflected shock is well captured by the present self-similar solution with $\alpha_1 = 0.693\,45$, slightly larger than Guderley's $\alpha^* = 0.688\,38$.

$$\frac{Z_3}{Z_2} = \frac{\tilde{p}_3}{F(\tilde{p}_3)}, \tag{3}$$

where

$$F(x) \equiv \frac{\gamma - 1 + (\gamma + 1)x}{\gamma + 1 + (\gamma - 1)x}, \tag{4}$$

denotes the Rankine–Hugoniot density compression ratio $F = \rho_{i+1}/\rho_i$ across the shock, expressed as a function of the pressure ratio $x = p_{i+1}/p_i$, where the subscripts $i$ and $i+1$ denote the upstream and downstream states, respectively.

Because of the singular behavior at the sonic line $Z = (U - \alpha^*)^2$, the trajectory originating from point 3 cannot smoothly cross this line with $\alpha = \alpha^*$, and thus no continuous solution exists connecting point 3 to $(U, Z) \to (0, 0)$ as $\xi \to \infty$ for this exponent. However, another eigenvalue, $\alpha = \alpha_1 = 0.693\,45$, yields a trajectory that smoothly connects point 3 to $(U, Z) = (0, 0)$ and crosses its own sonic line $Z = (U - \alpha_1)^2$, as shown by the red curve in Fig. 4(a). This exponent is only slightly ($\sim 0.7\%$) larger than $\alpha^*$.

The spatial profiles of density, velocity, and pressure in Figs. 4(b)–4(d) confirm that this locally self-similar solution closely matches the simulation results, supporting the validity of the present extension to the multi-shock configuration.

Figures 5(a) and 5(b) show the temporal evolutions of pressure and density obtained from the 1D hydrodynamic simulation corresponding to Fig. 2. The labels A–C and D–F indicate the converging and diverging phases, respectively. In the diverging phase, the two converging shocks merge into a single outgoing shock. The multi-stacked self-similar solution employs the exponent $\alpha_1 = 0.693\,45$, slightly larger than Guderley's value $\alpha^* = 0.688\,38$, and accurately captures the reflected shock structure.

## III. COMPRESSION SCALING

### A. Strong shock limits

Here, we estimate the maximum compression achievable after $N$-stacked shock waves collapse at the center and reflect as a single merged shock wave. As a starting point, we first consider the two-stacked case ($N = 2$) in planar geometry for simplicity, since it provides a transparent physical picture and serves as a bridge to the spherical case discussed later.

According to the Rankine–Hugoniot relations, the density compression ratios across each stage can be expressed in terms of the function $F(x)$ [see Eq. (4)] as follows: for the first shock, $\rho_1/\rho_0 = F(p_1/p_0)$; for the second shock, $\rho_2/\rho_1 = F(p_2/p_1)$; and for the reflected shock, $\rho_r/\rho_2 = F(p_r/p_2)$.

Thus, the overall density ratio $\tilde{\rho}_2 = \rho_2/\rho_0$ becomes a function of $\tilde{p}_1 = p_1/p_0$ and $\tilde{p}_2 = p_2/p_0$. To determine the optimal intermediate pressure that yields the highest total compression, we fix the final pressure $\tilde{p}_2$ and maximize $\tilde{\rho}_2$ with respect to $\tilde{p}_1$, i.e., $\partial \tilde{\rho}_2 / \partial \tilde{p}_1 = 0$, which gives the optimum condition $\tilde{p}_1 = \tilde{p}_2^{1/2}$. Substituting this result into the Rankine–Hugoniot expression yields the maximum value,

$$\frac{\rho_2}{\rho_0} = \left[F(\tilde{p}_2^{1/2})\right]^2 = \left[\frac{(\gamma+1)\tilde{p}_2^{1/2} + (\gamma-1)}{(\gamma-1)\tilde{p}_2^{1/2} + (\gamma+1)}\right]^2. \tag{5}$$

Proceeding in the same manner, the analysis can be extended to a general $N$-stacked configuration. For a fixed final pressure $\tilde{p}_N$, the condition for maximizing $\tilde{\rho}_N$ is obtained by solving $\partial \tilde{\rho}_N / \partial \tilde{p}_i = 0$ for $i = 1, 2, \ldots, N-1$. This yields an optimal geometric progression of intermediate pressures, $(\tilde{p}_1, \tilde{p}_2, \ldots, \tilde{p}_{N-1}) = (\tilde{p}_N^{1/N}, \tilde{p}_N^{2/N}, \ldots, \tilde{p}_N^{(N-1)/N})$, and the maximum compression ratio

$$\frac{\rho_N}{\rho_0} = \left[F(\tilde{p}_N^{1/N})\right]^N = \left[\frac{(\gamma+1)\tilde{p}_N^{1/N} + (\gamma-1)}{(\gamma-1)\tilde{p}_N^{1/N} + (\gamma+1)}\right]^N. \tag{6}$$

Next, we estimate the final compressed density $\rho_r$ after the reflection at the center. Consider two equivalent strong shocks, each characterized by the same compressed density $\rho_1$ and pressure $p_1$, propagating toward each other in an initially uniform medium of density $\rho_0$ and pressure $p_0 \ll p_1$. At the moment of collision, the two counter-propagating shocks form a stagnated region at the center, while reflected shocks propagate outward in opposite directions. The R–H relation, combined with the stagnation condition (i.e., vanishing fluid velocity at the reflection point), then yields the final density in the single-shock case ($N = 1$) as

$$\frac{\rho_r}{\rho_0} = g_0(\gamma) \equiv \frac{\gamma(\gamma+1)}{(\gamma-1)^2}. \quad (7)$$

For an ideal gas with $\gamma = 5/3$, this gives $\rho_r/\rho_0 = 10$. Equation (7) can be equivalently expressed using the strong-shock compression ratio $\rho_1/\rho_0 = F(\infty) = (\gamma+1)/(\gamma-1)$ as

$$\frac{\rho_r}{\rho_0} = \frac{\gamma}{\gamma-1} \cdot \frac{\rho_1}{\rho_0}. \quad (8)$$

In the case of $N$-stacked shocks, the overall maximum compression ratio after reflection at the center is obtained by replacing $\rho_1/\rho_0$ in Eq. (8) with $\rho_N/\rho_0$ given by Eq. (6) in the limit $\tilde{p}_N \to \infty$, since in the immediate vicinity of the center, the state produced by N-stacked shocks becomes locally indistinguishable from a single strong-shock-compressed state, leading to

$$\frac{\rho_r}{\rho_0} = g_0(\gamma)\left(\frac{\gamma+1}{\gamma-1}\right)^{N-1}, \quad \text{(planar)}, \quad (9)$$

which yields

$$\frac{\rho_r}{\rho_0} = 10 \times 4^{N-1}, \quad (\gamma = 5/3; \text{ planar}). \quad (10)$$

In spherical geometry, the optimal maximum density $\tilde{\rho}_r = \rho_r/\rho_0$, obtained after $N$-stacked converging shock waves coalesce infinitesimally close to the center and are subsequently reflected, can be derived in a manner analogous to the planar case. In the Guderley solution ($N = 1$) with $\gamma = 5/3$, for example, the density jump across the shock front is $\rho_{\xi=1}/\rho_0 = (\gamma+1)/(\gamma-1) = 4$, which gradually increases to $\tilde{\rho}_\infty^* \equiv \rho_{\xi=\infty}^*/\rho_0 = 9.55$. It should be noted that at the very moment of shock collapse at the center, the density profile in Guderley's self-similar solution becomes spatially uniform: $\rho_{r=r_a}^*/\rho_0 \to \tilde{\rho}_\infty^* = 9.55$ [see Fig. 6.17(b) in Ref. 3], just as in the planar case. Here, $r_a$ denotes an arbitrary fixed radius satisfying $R < r_a$. This uniformity arises because, in the self-similar solution, as the shock front converges to the center ($R(t) \to 0$), the terminal compression ratio $\tilde{\rho}_\infty^* = 9.55$ manifests more clearly at any fixed radius, due to $r_a/R \to \infty$. Thus, for $N = 1$ and $\gamma = 5/3$, the compression ratios in spherical geometry[2] are $\rho_1^*/\rho_0 = 9.55$ and $\rho_r^*/\rho_0 = 32.3$, in contrast to $\rho_1^*/\rho_0 = 4$ and $\rho_r^*/\rho_0 = 10$ in planar geometry. Consequently, the optimal compression ratio after the reflection of $N$-stacked shocks at the center is given by

$$\frac{\rho_r}{\rho_0} = 32.3 \times 9.55^{N-1}, \quad (\gamma = 5/3; \text{ spherical}). \quad (11)$$

This result represents the theoretical upper bound of compression achievable via multi-shock implosion in spherical geometry under ideal hydrodynamic conditions. It generalizes the classical Guderley solution to the $N$-shock regime and quantifies how sequential shock convergence and reflection can amplify the central density beyond the single-shock limit. In this sense, Eq. (11) serves as a benchmark for evaluating the ultimate performance of staged implosion schemes in high-energy-density physics.

Note that the factor $g_0(\gamma)$ appearing in Eqs. (7)–(9) represents the geometric effect inherent in the central shock reflection. The corresponding values are $g_0 = 10$ for planar geometry and $g_0 = 32.3$ for spherical geometry, the latter being obtained only numerically. This factor thus encapsulates the essential role of geometry in determining the ultimate compression at stagnation.

### B. Weak shock limits

We analyze the planar $N = 2$ case with a weak second shock, $P_2/P_1 = 1 + \epsilon$ ($\epsilon \ll 1$), and a strong first shock $P_1/P_0 \to \infty$. Using the Rankine–Hugoniot compression function $F(x)$ defined by Eq. (4) for the first, second, and reflected shocks, and imposing the same stagnation condition as in Sec. III A, a Taylor expansion to $O(\epsilon)$ gives

$$\frac{\rho_r}{\rho_0} \simeq g_0(\gamma) + g_1(\gamma)\,\epsilon, \quad (12)$$

with $g_0(\gamma)$ from Eq. (7) and $g_1(\gamma)$ as the first-order correction.

We absorb the weak-shock correction into a dimensionless exponent $\beta \equiv g_1/g_0$ (Table I), which represents the sensitivity of the final compression to an incremental increase in pressure, and rewrite using the single-shock reference $\tilde{\rho}_r^*$ in the form,

$$\frac{\rho_r}{\rho_0} = \tilde{\rho}_r^* \left(\frac{P_2}{P_1}\right)^\beta. \quad (13)$$

For an $N$-stacked sequence of weak shocks, assuming that each stage satisfies $P_{i+1}/P_i = 1 + O(\epsilon)$, the individual compression factors multiply, yielding

$$\frac{\rho_r}{\rho_0} = \tilde{\rho}_r^* \left(\frac{P_2}{P_1}\right)^\beta \left(\frac{P_3}{P_2}\right)^\beta \cdots \left(\frac{P_N}{P_{N-1}}\right)^\beta = \tilde{\rho}_r^* \left(\frac{P_N}{P_1}\right)^\beta. \quad (14)$$

Thus, within the weak-shock framework, $\rho_r/\rho_0$ depends only on the total pressure ratio $P_N/P_1$, and is independent of the intermediate pressure levels $\{P_2, P_3, \ldots, P_{N-1}\}$. As a specific case of Eq. (14), consider a pressure configuration in the form of a geometric sequence as derived in the strong-shock limit,

$$\frac{P_2}{P_1} = \frac{P_3}{P_2} = \cdots = \frac{P_N}{P_{N-1}} = \left(\frac{P_N}{P_1}\right)^{1/(N-1)} \equiv \hat{P}. \quad (15)$$

Then, the overall compression ratio reduces to

$$\frac{\rho_r}{\rho_0} = \tilde{\rho}_r^* \hat{P}^{\beta(N-1)}. \quad (16)$$

TABLE I. Dependence of the similarity exponent $\alpha^*$, the asymptotic density ratio $\tilde{\rho}_\infty^* = \rho_\infty^*/\rho_0$, the reflected density ratio $\tilde{\rho}_r^* = \rho_r^*/\rho_0$, the reflected pressure ratio $\tilde{p}_r^* = p_r^*/\rho_0 \dot{R}^2$, and the scaling exponent $\beta$ on the adiabatic index $\gamma$ for a single spherical shock ($N = 1$).

| $\gamma$ | $\alpha^*$ | $\tilde{\rho}_\infty^*$ | $\tilde{\rho}_r^*$ | $\tilde{p}_r^*$ | $\beta$ |
|---|---|---|---|---|---|
| 5/3 | 0.688 38 | 9.5495 | 32.292 | 23.007 | 0.723 61 |
| 1.60 | 0.694 18 | 11.066 | 42.670 | 33.112 | 0.754 22 |
| 1.55 | 0.699 05 | 12.527 | 54.339 | 45.342 | 0.778 75 |
| 1.50 | 0.704 43 | 14.387 | 71.744 | 64.971 | 0.804 71 |
| 1.45 | 0.710 42 | 16.812 | 99.094 | 98.529 | 0.832 19 |

For $\gamma = 5/3$, we have $\tilde{\rho}_r^* = 10$ (planar) and $\tilde{\rho}_r^* = 32.3$ (spherical), with the common exponent $\beta = 0.724$, which yields the final scaling law

$$\frac{\rho_r}{\rho_0} = \begin{cases} 10\,\hat{P}^{0.724(N-1)}, & \text{(planar)} \\ 32.3\,\hat{P}^{0.724(N-1)}, & \text{(spherical)} \end{cases} \quad \text{for} \quad \gamma = \frac{5}{3}. \quad (17)$$

Correspondingly, for $\gamma = 5/3$, the critical value of the stage pressure ratio $\hat{P}_c$, which marks the boundary between the weak- and strong-shock regimes, can be estimated from Eqs. (10), (11), and (17), using $\tilde{\rho}_\infty^* = 4$ for planar geometry and $\tilde{\rho}_\infty^* = 9.55$ for spherical geometry, as

$$\hat{P}_c = (\tilde{\rho}_\infty^*)^{1/\beta} = \begin{cases} 6.8, & \text{(planar)} \\ 22.6, & \text{(spherical)} \end{cases} \quad \text{for} \quad \gamma = \frac{5}{3}. \quad (18)$$

Table I summarizes the numerically obtained parameters for spherical geometry, showing how the similarity exponent $\alpha^*$, the asymptotic density $\tilde{\rho}_\infty^*$, the reflected density $\tilde{\rho}_r^*$, and the scaling exponent $\beta$ depend on the adiabatic index $\gamma$. A clear and consistent trend is observed: both $\tilde{\rho}_r^*$ and $\beta$ increase monotonically with decreasing $\gamma$, indicating that softer equations of state yield higher final compression ratios [compare Eq. (16)]. Physically, as $\gamma \to 1$, the fluid gains more internal degrees of freedom, becoming increasingly compressible. Radiative losses or high-$Z$ doping can effectively lower the local adiabatic index, mitigating stagnation heating and facilitating higher-density compression in converging shock flows.

### C. Areal mass density

The areal mass density (or areal density $\rho R$) of the reflected shock $\langle \rho r \rangle_r \equiv \int_0^{R(t)} \rho(t, r)\,dr$, is a key parameter in determining the fuel burn fraction in ICF. The self-similar solution gives the reflected-shock density profile in the approximate form $\rho(r,t)/\rho_r \simeq (r/R(t))^{0.7}$ for $r \leq R(t)$. Assuming the expanding front at $r = R(t)$ originates from a fluid element initially located at $r_0$, the mass conservation condition, $\frac{4\pi}{3}\rho_0 r_0^3 = \int_0^{R(t)} 4\pi r^2 \rho(r,t)\,dr$, yields the approximate relation

$$\frac{\langle \rho r \rangle_r}{\rho_0 r_0} \simeq 0.63 \left(\frac{\rho_r}{\rho_0}\right)^{2/3} \simeq 20.3\,\hat{P}^{0.483(N-1)}, \quad (19)$$

where the numerical coefficient 0.63 arises from the assumed similarity profile, and the second equality follows from Eq. (17). This relation holds for both the single-shock limit (N = 1, corresponding to the Guderley solution) and multi-shock cases, while the achievable areal density increases rapidly with N. For instance, taking $\langle \rho r \rangle_r = 1 - 2$ g/cm$^2$ as the requirement for achieving medium burn fraction, $r_0 \approx 400 - 800\,\mu$m assuming $\rho_r/\rho_0 = 2000$ and $\rho_0 = 0.25$ g/cm$^3$ for solid DT. This scaling clearly demonstrates that ignition-relevant areal densities can be attained with practical target dimensions, underscoring the feasibility of multi-shock-driven designs.

### D. Fluid simulations and comparison with theory

For convenience, the analytical results derived in the preceding sections [Eqs. (11), (17), and (18)] can be summarized in a unified, piecewise form for the representative case,

$$\frac{\rho_r}{\rho_0} = \begin{cases} 32.3\,\hat{P}^{0.724(N-1)}, & (\hat{P} \lesssim 23), \\ 32.3 \times 9.55^{N-1}, & (\hat{P} \gtrsim 23), \end{cases} \quad (\gamma = 5/3; \text{ spherical}), \quad (20)$$

where $\hat{P}_c \simeq 23$ denotes the critical stage pressure ratio separating the weak- and strong-shock regimes. This compact form bridges the analytical limits obtained earlier and serves as a convenient benchmark for numerical validation.

To test the analytical predictions in Eq. (20), one-dimensional Lagrangian hydrodynamic simulations were carried out for an ideal gas with $\gamma = 5/3$. Calculations were performed in both planar and spherical geometries for $N$-stacked configurations with $N = 1 - 5$. The stage pressure ratios were taken as $\hat{P} = 2^{j/2}$ with integers $j \leq 12$, forming the geometric progression defined in Eq. (15).

Figure 6 shows the results for planar geometry, where the final compression ratio $\rho_r/\rho_0$ is plotted as a function of $\hat{P}$. The numerical data are in excellent agreement with the analytical predictions given by Eqs. (10) and (17), corresponding to the strong- and weak-shock limits. Thin solid curves denote numerical solutions of the full Rankine–Hugoniot relations without limiting approximations. Notably, the weak-shock scaling law [Eq. (17)] reproduces the simulations with high accuracy not only in the weakly nonlinear regime ($\hat{P} - 1 \ll 1$) but also in the moderately nonlinear domain ($\hat{P} \sim 1 - 7$), indicating a wider range of validity than formally expected.

Figure 7 presents the corresponding results for spherical geometry. Here the agreement with the analytical model [Eq. (20)] is even more striking. The weak-shock scaling law remains accurate over a broad range up to $\hat{P} \approx \hat{P}_c \sim 23$, despite the strongly nonlinear character of the flow.

This extended validity suggests that the perturbative scaling law captures the essential compression dynamics well beyond its nominal regime. A plausible interpretation is that geometric convergence in spherical geometry introduces a universal amplification mechanism that stabilizes the flow. Such geometric focusing enhances compression while smoothing short-wavelength perturbations, consistent with the cutoff mode behavior found in the two-dimensional perturbation analysis of converging shocks by Murakami *et al.*[21]

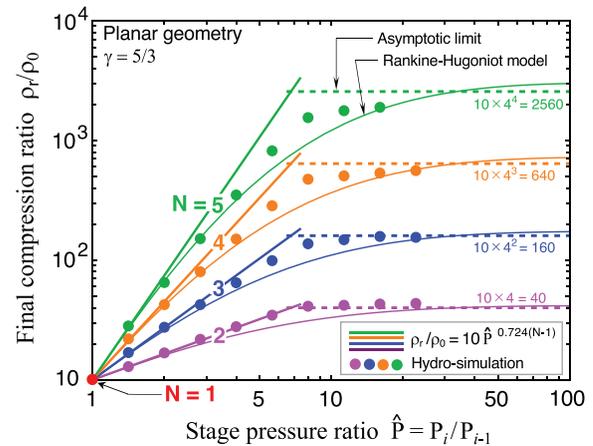

FIG. 6. Final compression ratio $\rho_r/\rho_0$ as a function of the stage pressure ratio $\hat{P} \equiv P_{i+1}/P_i$ for $N = 2 - 5$ stacked shocks in planar geometry, based on the Rankine–Hugoniot (R–H) relations under the strong-shock approximation for $\gamma = 5/3$. Each curve corresponds to a fixed number of shocks $N$, and follows the asymptotic scaling law $\rho_r/\rho_0 \simeq 10 \times 4^{N-1}$ [Eq. (10)], as indicated by the horizontal dashed lines. Simulation results (colored dots) are overlaid and exhibit excellent agreement with the theoretical scaling law $\rho_r/\rho_0 \simeq 10\,\hat{P}^{0.724(N-1)}$ [Eq. (20)], confirming the validity of the scaling behavior in the planar case.

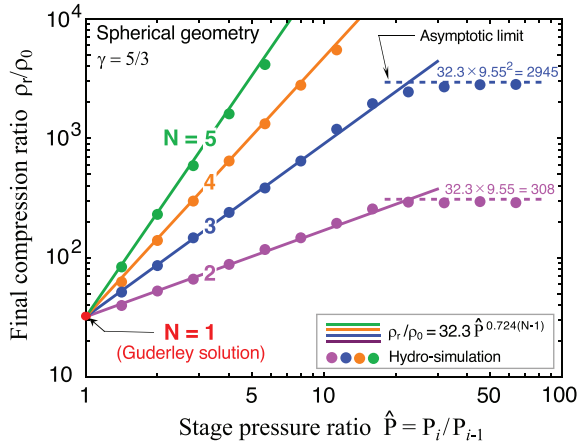

**FIG. 7.** Final compression ratio $\rho_r/\rho_0$ as a function of the stage pressure ratio $\hat{P} \equiv P_{i+1}/P_i$ for $N = 1 - 5$ stacked shocks in spherical geometry. The analytical prediction $\rho_r/\rho_0 \simeq 32.3 \hat{P}^{0.724(N-1)}$ [Eq. (20); $\gamma = 5/3$] shows excellent agreement with the hydrodynamic simulations (colored dots), even in the asymptotic limit, demonstrating the robustness of the multi-shock compression scaling. Reproduced with permission from Murakami, Phys. Rev. E **112**, 055206 (2025) under CC BY 4.0.[20]

This self-regularizing tendency, arising purely from focusing geometry, may reflect a deeper universality in converging flows. If confirmed, it would reinforce the predictive power and practical relevance of self-similar models in high-energy-density compression dynamics.

### E. Dynamic equivalence between pressure and density in the stagnated core

As discussed above, geometric convergence of shocks acts as a natural amplification mechanism that regularizes nonlinear effects and promotes self-regulation in the implosion. This behavior is further reflected in a dynamic equivalence between pressure and density within the stagnated core.

Within the self-similar framework, the reflected pressure is expected to remain nearly uniform in the post-reflection quasi-isobaric region and to follow a scaling relation analogous to that of density,

$$\tilde{p}_r \equiv \frac{p_r}{\rho_0 \dot{R}^2} = \tilde{p}_r^* \hat{P}^{\beta(N-1)}, \quad (21)$$

where $p_r$ denotes the pressure immediately after shock reflection. As listed in Table I, the reference coefficient is $\tilde{p}_r^* = 23.0$ for $\gamma = 5/3$. Figure 8 compares this theoretical scaling with hydrodynamic simulations and shows excellent agreement. Notably, Figs. 7 and 8 indicate that both $\rho_r$ and $p_r$ exhibit nearly identical dependencies on the stage pressure ratio $\hat{P}$, despite arising from different aspects of the implosion dynamics.

This near coincidence follows directly from radial momentum conservation. At the instant of reflection, the cumulative impulse of the converging shocks is balanced by the internal pressure of the compressed core. Since the mass is already highly concentrated, the quasi-equilibrium pressure scales as $p_r \sim \rho_r v^2 \sim \rho_r \dot{R}^2$, leading naturally to the same functional dependence on $\hat{P}$ for both quantities. This dynamic equivalence highlights the self-regulating nature of the implosion, in which the total energy input simultaneously determines both the attainable compression and the internal pressure.

The correspondence is therefore not merely numerical but provides a useful diagnostic linkage. In experiments, measurement of one quantity (e.g., pressure) can be used to infer the other (e.g., density), offering an internal consistency check for theory and simulation. Such coherence further supports the stacked-shock concept as a unified and predictive framework for achieving ultrahigh compression.

### F. Physical significance of off-center shock reflection

An important structural feature revealed by the present multi-shock implosion model is that the final maximum compression is not reached at the geometric center ($r = 0$), but at a finite radius ($r = r_{\min}$), where the innermost converging shock is dynamically reflected within the highly compressed core. We refer to this phenomenon as *off-center shock reflection*.

Although the analytical scaling laws [Eq. (20)] are derived within a self-similar framework extending the classical Guderley-type solution, they remain quantitatively valid even though the density peak appears away from the center. In the Guderley solution, only pressure and temperature exhibit singular growth at $r = 0$, while the density remains finite. In realistic hydrodynamics, the rapid increase in pressure, temperature, and flow gradients near the center leads to the breakdown of the ideal description, with kinetic and non-collisional effects becoming important.

The emergence of a finite reflection radius $r_{\min}$ is therefore essential. It marks the location where the inward-propagating shock is halted by the steep pressure rise of the precompressed core, so that the ram pressure of the inflow balances the internal pressure, $\rho u^2 \sim p$.

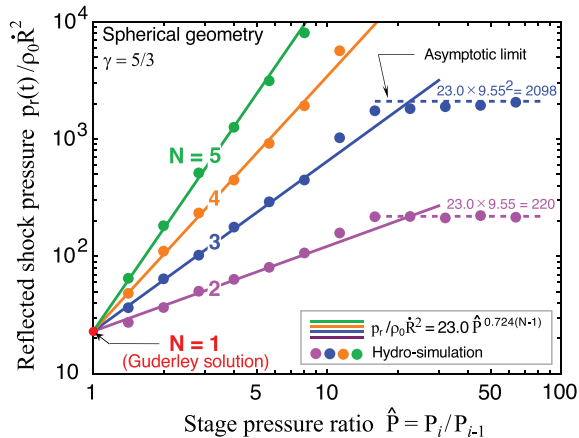

**FIG. 8.** The reflected shock pressure $p_r(t)/\rho_0 \dot{R}^2$ as a function of the stage pressure ratio $\hat{P} \equiv P_{i+1}/P_i$ for $N = 1 - 5$ stacked shocks in spherical geometry. The analytical estimates follow the asymptotic expression $p_r(t)/\rho_0 \dot{R}^2 \simeq 23.0 \times 9.55^{N-1}$, while the simulation results (colored dots) are overlaid and closely follow the theoretical scaling law $p_r(t)/\rho_0 \dot{R}^2 \simeq 23.0 \hat{P}^{0.724(N-1)}$ [Eq. (21)] for $\gamma = 5/3$, demonstrating the robustness of the multi-shock compression scaling.

This balance defines the reflection interface, typically two to three orders of magnitude smaller than the initial target radius ($r_{\min}/R_0 \sim 10^{-2}-10^{-3}$ for representative parameters). At this point, the stacked-shock compression terminates, and an outward-propagating reflected shock is launched, producing the maximum density at a finite radius.

This off-center reflection provides a natural self-regulating mechanism that preserves the self-similar structure while enforcing physical regularity. The solution remains scale-invariant in the converging region down to $r_{\min}$, and then transitions smoothly to an outward-propagating reflected shock. Thus, ultrahigh compression can be achieved without singular collapse, with $r_{\min}$ acting as the natural end point of cumulative shock compression. Geometric focusing should therefore be viewed as a stable finite-radius process, rather than a true central singularity.

### G. Robustness against timing mismatch

To assess how sensitive the optimized compression scheme is to deviations in the timing of the driving pressure pulses, we carried out a series of simulations for $N = 3$ and 4. Figure 9 shows how the final compression ratio decreases as the pulse intervals deviate from their optimal values. Here, the timing offset of each pulse is denoted by $\Delta t_i$; for example, positive and negative values of $\Delta t_2$ correspond to a delay or an advance, respectively, of the second pressure pulse $P_2$ relative to its optimal timing (compare Fig. 1). Although substantial degradation occurs for large timing errors, the results indicate that moderate mistiming can be tolerated without a significant loss in compression performance. A rough estimate from Fig. 9 further suggests that the final compression remains largely intact as long as the timing mismatch is kept within roughly 50–100 ps for the representative cases considered here.

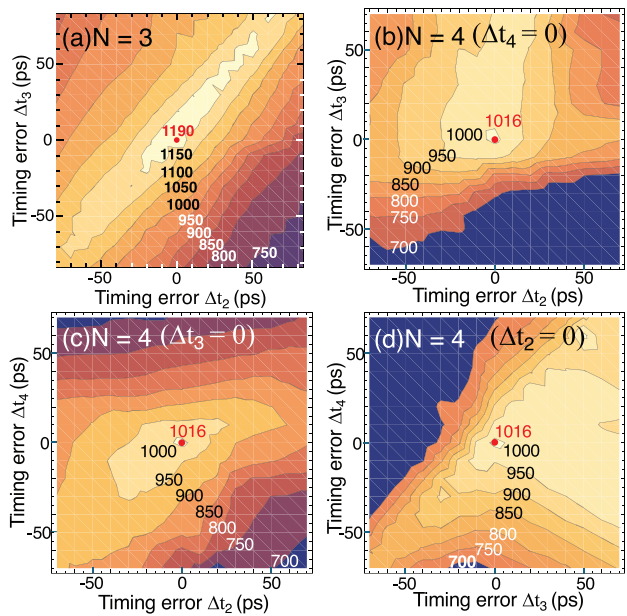

**FIG. 9.** Contour plots showing the sensitivity of the maximum compression $\rho_r$ to relative timing errors among multiple shocks for $N = 3$ and 4. Panels (a)–(d) correspond to different fixed-delay configurations. Each axis denotes the timing error (ps) between neighboring shocks, and the color map indicates the maximum compression ratio $\rho_r/\rho_0$. A narrow window of high-performance near-perfect synchronization reveals a degree of robustness relevant to practical implementation.

This tolerance to synchronization errors reflects a key advantage of the multi-shock approach. Unlike highly resonant or phase-sensitive systems, where performance critically depends on precise phase matching, the multi-shock scheme sustains effective compression even when the pulse timing is not perfectly synchronized, thus relaxing the requirements for precision in experimental implementations. Such robustness is particularly valuable in practical settings where accurate pulse shaping and synchronization are technically demanding. This behavior reflects the cumulative nature of multi-shock compression, in which the final state is governed primarily by the integrated impulse rather than by the exact arrival time of individual shocks.

Further insight into this robustness comes from the density response under mistimed conditions. If the prescribed pressure sequence $\{P_1, P_2, \ldots, P_N\}$ is preserved and successive shocks reach the center within an acceptable delay, the compression remains largely preserved. In contrast, once a later shock overtakes an earlier one still converging inward, the density-compression performance drops sharply: such overtaking generates outward rarefaction waves at the collision point, disrupting the coherent shock coalescence that sustains cumulative compression. This sensitivity to the relative timing of successive shocks is consistent with previous analyses of shock-timing effects in laser-driven implosions.[16]

### H. Entropy reduction and the isentropic limit in multi-shock compression

Finally, from a thermodynamic viewpoint, the robustness and self-regulation discussed above can be interpreted as manifestations of entropy suppression inherent in multi-shock compression. A quantitative measure of irreversibility in each shock stage can be obtained from the weak-shock expansion of the R–H relations for a perfect gas. For a small pressure increment $\epsilon = (P_2 - P_1)/P_1 \ll 1$, the entropy increase per unit mass is[2]

$$\Delta s = c_v \frac{(\gamma + 1)^2}{12 \gamma^2} \epsilon^3 + O(\epsilon^4), \quad (22)$$

where $c_v$ is the specific heat at constant volume. The first and second-order terms cancel exactly, so the entropy jump appears only at third order, implying that splitting a large compression into many weak shocks strongly suppresses entropy production.

With the total pressure ratio $\mathcal{P}_{\text{total}} = P_N/P_1$ under $N-1$ identical weak shocks, each pressure increment for $i^{\text{th}}$ shock turns out to be $\epsilon_i = (P_{i+1}/P_i) - 1 \simeq (\ln \mathcal{P}_{\text{total}})/(N-1)$. Summing Eq. (22) over $N-1$ stages gives

$$\Delta s_{\text{total}} = C(\gamma) \frac{(\ln \mathcal{P}_{\text{total}})^3}{(N-1)^2}, \quad C(\gamma) = c_v \frac{(\gamma + 1)^2}{12 \gamma^2}. \quad (23)$$

For $\gamma = 5/3$, $C(\gamma)/K \simeq 0.32$, so that $\Delta s_{\text{total}}/K = 0.32 (\ln \mathcal{P}_{\text{total}})^3/(N-1)^2$, where $K = c_p - c_v$ is the specific gas constant. This $(N-1)^{-2} \simeq N^{-2}$ scaling quantifies how the cumulative entropy production diminishes as $N \to \infty$. Thus, entropy

## IV. DISCUSSION AND OUTLOOK

### A. Structural advantages of volumetric compression over shell designs

As discussed in Sec. III E, the derived scaling law, Eq. (20), reveals a robust relation between the final compression ratio $\rho_r/\rho_0$ and the stage pressure ratio $\hat{P}$. This robustness originates from a geometric amplification inherent in spherical convergence, which regularizes nonlinear effects and enables stable, volumetric energy delivery.

In contrast to conventional shell-based implosions, the present scheme proceeds via volumetric compression and does not rely on a discrete material interface separating fluids of different densities. As a result, the effective local Atwood number remains small throughout the implosion. In addition, the deceleration phase associated with shock reflection is extremely short in duration, further limiting the time available for instability amplification even if locally seeded. Consequently, while small-amplitude perturbations may exist, Rayleigh–Taylor growth is expected to remain strongly mitigated in the parameter regime considered here.

A related concern is the possible development of Richtmyer–Meshkov (RM)-type perturbations due to repeated shock passages. In the present configuration, however, the same physical conditions that mitigate Rayleigh–Taylor growth are also expected to limit sustained RM amplification. As a result, although RM perturbations may be seeded, they are unlikely to grow to a level that significantly compromises the overall robustness of the multi-shock implosion.

Although a uniform solid sphere may appear counterintuitive to those familiar with shell targets, our results demonstrate that under optimally stacked converging shocks, solid targets can achieve efficient volumetric compression, attaining or even exceeding the density gains of hollow-shell implosions. This finding suggests that conventional design intuitions in inertial fusion should be reconsidered within this theoretical framework.

While the preceding discussion focused on density amplification, it is instructive to contrast the present uniform-sphere model with thin-shell configurations more commonly used in ICF. Kemp and Meyer-ter-Vehn[24] derived a cubic scaling law, $p_s \propto M_0^3$, for the stagnation pressure in thin-shell implosions. However, that law presupposes a finite stagnation region where the reflected shock reaches a quasi-steady plateau. In contrast, the present model describes a uniform sphere whose converging shock collapses at a finite radius $r_{\min}$ without forming a stable cavity. At this point, the density remains finite ($\rho \rightarrow \rho_{\max} < \infty$) while both pressure and temperature diverge ($p, T \rightarrow \infty$). Accordingly, a finite "stagnation pressure" cannot be defined; the physically meaningful quantities are the *final reflected density* $\rho_r$ and the *post-reflection plateau pressure* evaluated at a small but finite radius $r_{\mathrm{ref}} > r_{\min}$. In this context, the cubic scaling $p_s \propto M_0^3$ is not directly applicable to the present uniform-sphere geometry. Rather, the scaling relations derived here provide a complementary and more general framework for describing stagnation-free implosions driven by volumetric shock convergence.

Beyond these structural differences, the volumetric nature of the multi-shock implosion also has important practical implications for laser-driven implementations. In particular, the staged compression process introduces an additional degree of freedom through the number of shocks $N$, which plays a central role in determining both the achievable entropy reduction and the required laser dynamic range. This aspect is examined in the following subsection.

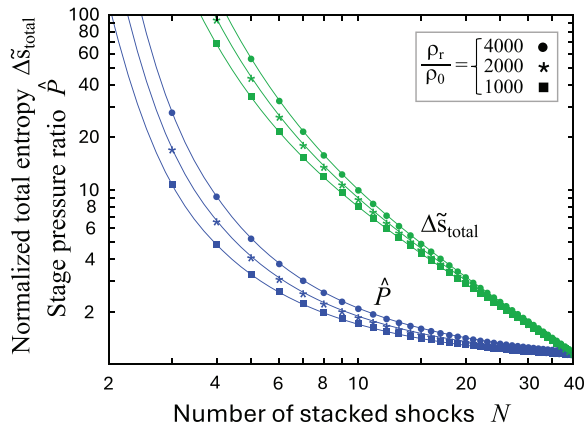

**FIG. 10.** Consequences of stacking a finite number of converging shocks at a fixed target reflected-shock compression. Shown are the normalized cumulative entropy production $\Delta\tilde{s}_{\mathrm{total}} \equiv \Delta s_{\mathrm{total}}/K$ (green curves), and the stage pressure ratio $\hat{P} \equiv P_{i+1}/P_i$ (blue curves) as functions of the number of shocks $N$. The entropy production decreases monotonically with increasing $N$, approaching the quasi-isentropic limit as $N \to \infty$. For each quantity, the three curves correspond to different final compression levels achieved at the reflected shock.

generation progressively decreases with increasing $N$, approaching the quasi-isentropic limit.

Figure 10 illustrates the dependence of the normalized cumulative entropy production and the stage pressure ratio on the number of shocks $N$ for a fixed overall compression level. The normalized entropy increase, $\Delta\tilde{s}_{\mathrm{total}} \equiv \Delta s_{\mathrm{total}}/K$, decreases monotonically as $N$ increases, indicating that the compression process approaches the quasi-isentropic limit in the multi-stage regime. At the same time, the required stage pressure ratio $\hat{P}$ approaches unity, reflecting a smoother redistribution of the compression work over successive shock transitions. For each quantity, the three curves correspond to different final compression levels achieved at the reflected shock, as indicated in the inset. This figure provides a compact visualization of how staged converging shocks reduce entropy production and gradually bridge the gap between strong-shock compression and the ideal isentropic limit.

This limit conceptually connects to the classical isentropic compression analysis by Kidder,[22] who studied isentropic compression of spherically symmetric targets with spatially tailored initial density and temporally tailored pressure profiles. More recently, Nagatomo *et al.*[23] performed radiation-hydrodynamic simulations of such multi-step pulse compression in the context of fast-ignition target design.

While their study demonstrated the feasibility and robustness of tailored step-pulse sequences for quasi-isentropic compression, the present work develops an analytic self-similar framework that generalizes this concept to an arbitrary number of stacked converging shocks, providing a theoretical foundation for its scaling and limiting behavior.

It should be stressed, however, that the quasi-isentropic limit reached as $N \to \infty$ in the present framework is fundamentally different from the homogeneous isentropic compression described by the Kidder model. Specifically, the present implosion remains constrained by inward-propagating compressive characteristics and their finite-time causality; see Appendix A.

## B. Finite-N optimization under realistic laser constraints

An important aspect of the present scaling framework concerns the practical realization of multi-shock implosions under realistic laser-driven conditions. In contrast to idealized continuous-drive scenarios, the staged nature of the compression introduces a discrete control parameter, the number of shocks $N$, which influences both entropy generation and the robustness of the compression dynamics.

In realistic laser-driven implosions, even a moderately structured (one-humped) pressure profile tends to steepen naturally into shock-like fronts. Once formed, the subsequent dynamics are governed primarily by the sequence of shock jumps rather than by the detailed temporal history of the driving waveform. This behavior reflects the causal nature of shock propagation: the compression at each stage is determined by local jump conditions and characteristic propagation, which progressively erase fine temporal structure in the initial drive, while preserving the causal ordering of shock arrivals.

As a result, a finite sequence of discrete shocks can reproduce, to a large extent, the same volumetric compression state that might otherwise be associated with an idealized continuous ramp.

This has an important practical implication. Instead of requiring precise control of the drive profile over the entire pulse duration, the essential requirement becomes the proper timing and ordering of a finite number of shock generations. From an engineering perspective, this substantially relaxes the need for highly fine-tuned, continuously shaped pulses, while still allowing the formation of a strongly compressed core.

From a physical and technological standpoint, laser-driven ablation operates within a finite intensity window. Efficient plasma formation and stable ablation typically require laser intensities of order $I_L \gtrsim (2-5) \times 10^{11}$ W/cm$^2$, while at higher intensities, $I_L \lambda_L^2 \gtrsim 10^{15}$ W cm$^{-2}$ $\mu$m$^2$, parametric instabilities and hot-electron generation become increasingly significant. Consequently, practical compression experiments are naturally carried out within a bounded range of usable drive conditions. Within this regime, staged compression by a finite number of shocks is physically well matched to the accessible parameter space and provides a robust route to achieving high volumetric compression without relying on extremely fine temporal control over a continuous drive.

In this sense, the finite-$N$ configuration should not be viewed merely as an approximation to the formal isentropic limit $N \to \infty$. Rather, it represents a practically favorable regime in which the causal structure of shock propagation, the natural steepening of pressure profiles, and the finite dynamic range of laser-driven ablation together support stable and efficient compression. This interpretation is consistent with the broader distinction, discussed in Appendix A, between characteristic-controlled implosion dynamics and quasi-static compression models.

We note that the reflected-shock phase formally satisfies the Rayleigh–Taylor sign condition; however, owing to the volumetric nature of the compression and the short effective deceleration time, the resulting instability remains transient and does not evolve into a destructive thin-shell-type growth.

## C. Dissipative effects and future extensions

While the present analysis focuses on idealized adiabatic shocks, Murakami et al.[13] demonstrated that thermal conduction during stagnation regularizes the otherwise divergent central temperature. Balancing compressive heating, $p\nabla \cdot \mathbf{u} \sim p_s/\tau$, with conductive losses, $\nabla \cdot [\kappa(T)\nabla T] \sim \kappa_0 T_c^{7/2}/L^2$, and adopting the hydrodynamic scale $L \sim c_s \tau$, one obtains the scaling

$$T_c \sim \left(\frac{p_s^2 \tau}{\rho \kappa_0}\right)^{2/7}.$$

The exponent $2/7$ originates from the $T^{5/2}$ dependence of the Spitzer conductivity and provides a geometry-independent estimate for the finite central temperature in diffusive stagnation. Physically, this represents a universal diffusive regularization of the central singularity—the conversion of an unbounded adiabatic spike into a finite, self-limited peak governed by transport processes.

This framework elucidates how dissipative mechanisms constrain the achievable compression and inform the design of targets less sensitive to pulse shaping and instabilities. Future extensions will incorporate radiation transport and multidimensional effects to evaluate these mechanisms in realistic high-energy-density environments. Ultimately, this framework establishes a unified theoretical basis that connects ideal self-similar solutions with dissipative implosions, thereby bridging analytic theory, numerical modeling, and experimental design in the study of high-energy-density matter. Furthermore, the present framework may also provide a useful basis for assessing alternative fusion concepts that rely on ultrahigh compression of solid-density fuel, including proton–boron schemes.[25]

A systematic extension of the present scaling theory to include dissipative effects, such as thermal conduction, and their impact on the maximum achievable compression and pressure, will be addressed in future work.

## V. SUMMARY

We have developed a class of self-similar solutions that generalize the classical Guderley framework to describe multi-shock implosions of solid-density targets. The analytic theory yields explicit scaling laws for the final compression of the form $\rho_r/\rho_0 \propto \hat{P}^{\beta(N-1)}$, where $N$ is the number of shocks, $\hat{P}$ the stage pressure ratio, and $\beta$ depends on the adiabatic index $\gamma$. These scaling relations have been validated by one-dimensional hydrodynamic simulations across a wide parameter range, extending from the weak- to strong-shock regimes, and approach a quasi-isentropic character as $N$ increases.

A central result of this work is that the maximum compression is realized not at the geometric center but through *off-center shock reflection*. This mechanism provides a natural cutoff to the otherwise singular behavior of the classical Guderley solution while preserving self-similar scaling. In this sense, off-center reflection emerges as the key physical process that regularizes the implosion and makes the derived scaling laws physically realizable.

A further advantage of this volumetric multi-shock scheme lies in its strong resilience against Rayleigh–Taylor instabilities, in contrast to conventional thin-shell implosions. Together with its analytic tractability and geometric robustness, this framework offers predictive guidance for the design of next-generation, largely instability-resistant compression schemes in high-energy-density physics and inertial confinement fusion. We finally note that such interface-driven perturbations, including Richtmyer–Meshkov modes, while not strictly absent, are expected to remain self-limited under the present volumetric and short-deceleration conditions, as discussed in Sec. IV A.

Looking ahead, future extensions should incorporate additional physical effects—such as radiation transport, thermal conduction, realistic equations of state, and multidimensional stability—to assess the limits of self-similarity under dissipative and nonideal conditions. Collectively, the present results establish a coherent and physically grounded theoretical framework that connects classical self-similar implosion theory with finite, physically realizable multi-shock compression and provides a predictive basis for exploring ultrahigh-compression regimes of matter.

Finally, we hope that the present results contribute, even in a modest way, to the long-standing search for physically robust pathways toward ultrahigh-density compression—a central goal that the high-energy-density and inertial-fusion communities have pursued for many decades.

## SUPPLEMENTARY MATERIAL

See the supplementary material for a statement on figure reuse, indicating that Figs. 2–4 are adapted from the author's previous publication under a CC BY 4.0 license.

## ACKNOWLEDGMENTS

This work was supported by the Japan Society for the Promotion of Science (JSPS) and the Kansai Electric Power Company, Incorporated (KEPCO). The author thanks Professor J. Meyer-ter-Vehn for insightful discussions that guided the theory and Dr. A. Velikovich for valuable comments and for providing key references that improved the manuscript.

## AUTHOR DECLARATIONS
### Conflict of Interest

The author has no conflicts to disclose.

### Author Contributions

**M. Murakami:** Conceptualization (lead).

## DATA AVAILABILITY

The data that support the findings of this study are available from the corresponding author upon reasonable request.

## APPENDIX A: INCOMPATIBILITY OF THE KIDDER MODEL WITH THE LIMIT $N \to \infty$ IN SPHERICAL IMPLOSIONS

The Kidder model describes quasi-static, homogeneous isentropic compression, in which the temporal evolution is prescribed without enforcing causal transmission from the boundary to the center.[22] In contrast, the implosion studied here is governed by inward-propagating compressive characteristics that transmit boundary information to the center within a finite time. As a result, the boundary evolution $R(t)$ and pressure $p(t)$ determined by characteristic-controlled dynamics are fundamentally incompatible with the Kidder solution, even in the formal limit $N \to \infty$. The quasi-isentropic limit obtained in the present multi-shock framework, therefore, cannot be reduced to homogeneous isentropic compression of the Kidder type.

### 1. Governing equations and differential operators

We consider a spherically symmetric, inviscid, isentropic flow $p = p(\rho)$, with sound speed

$$c^2 \equiv \frac{dp}{d\rho}. \tag{A1}$$

Two derivatives must be distinguished. The material derivative is

$$\frac{D}{Dt} \equiv \partial_t + u\,\partial_r, \tag{A2}$$

and the derivative along the inward characteristic $C^-$ ($dr/dt = u - c$) is

$$D_- \equiv \partial_t + (u - c)\,\partial_r. \tag{A3}$$

They are related by

$$D_- = \frac{D}{Dt} - c\,\partial_r. \tag{A4}$$

### 2. Euler equations in spherical symmetry

The Euler equations can be written as

$$\frac{D\rho}{Dt} = -\rho\left(\partial_r u + \frac{2u}{r}\right), \tag{A5}$$

$$\frac{Du}{Dt} = -\frac{1}{\rho}\partial_r p = -\frac{c^2}{\rho}\,\partial_r \rho. \tag{A6}$$

For a polytropic isentrope $p = K\rho^\gamma$,

$$\frac{Dc}{Dt} = -\frac{\gamma - 1}{2}\,c\left(\partial_r u + \frac{2u}{r}\right). \tag{A7}$$

### 3. Breakdown of Riemann invariance in spherical geometry

In planar one-dimensional isentropic flow, the Riemann invariant

$$J_- \equiv u - \frac{2c}{\gamma - 1}, \tag{A8}$$

is conserved along the inward characteristic.[2] In spherical geometry, this invariance is broken by a geometric source term. Taking the $C^-$ derivative and using Eq. (A4) gives

$$D_- J_- = \frac{2cu}{r}. \tag{A9}$$

Thus, spherical convergence introduces an intrinsic $1/r$ contribution that continuously modifies the inward characteristic, an effect absent in planar flow and not captured by quasi-static models such as Kidder's.

### 4. Finite-time causal transmission from boundary to center

Let $t_b$ denote the time at which the boundary state is specified at $r = R(t_b)$. The inward characteristic defined by $dr/dt = u - c$ reaches the center at a finite time $t_c(t_b)$ and satisfying $r(t_c) = 0$.

This establishes a causal delay between boundary forcing and central response:

$$t_c(t_b) - t_b = \int_0^{R(t_b)} \frac{dr}{c(r,t) - u(r,t)}. \quad (A10)$$

### 5. Explicit contrast of boundary evolution for $\gamma = 5/3$

For $\gamma = 5/3$, the two descriptions yield qualitatively different boundary histories.

*Kidder model (homogeneous isentropic compression).* A representative homologous solution is

$$R_K(t) = R_0\sqrt{1-\tau^2}, \quad \tau \equiv t/t_c. \quad (A11)$$

Uniform compression implies $\rho \propto R_K^{-3}$ and

$$p_K(t) \propto \rho^\gamma \propto R_K(t)^{-3\gamma} \propto (1-\tau^2)^{-5/2}. \quad (A12)$$

The essential feature is that the temporal evolution is prescribed without an explicit characteristic constraint.

*Characteristic-controlled implosion* ($N \to \infty$). In the present formulation, the boundary motion is constrained by the arrival of inward compressive characteristics, leading to

$$R(t) \propto (1-\tau)^\nu. \quad (A13)$$

Introducing a characteristic density scale $\bar{\rho} \sim M/R^3$ and $p \propto \rho^\gamma$ gives

$$p(t) \propto R(t)^{-3\gamma} \propto (1-\tau)^{-5\nu}. \quad (A14)$$

The exponent $\nu$ is set by causal characteristic dynamics in spherical geometry and is unrelated to the Guderley shock exponent.